# “What If I Had Participated?”—A Machine Learning-Based Counterfactual Analysis of Non-Participants’ Success Potential in STEM Competitions: The Case of the German Physics Olympiad

Paul Tschisgale[1*], Knut Neumann[1]

[1] *Leibniz Institute for Science and Mathematics Education (IPN), Kiel, Germany*

Correspondence: tschisgale@leibniz-ipn.de

ABSTRACT

STEM student competitions are widely recognized as valuable enrichment activities that can positively influence participants’ academic and career trajectories. They aim to identify capable, science-interested students and further support them in developing their STEM-related abilities and interests. When such students do not participate, they may appear to be missed not only by the competition but by this form of support altogether—unless they are reached through other competitions. The present study therefore investigated to what extent students who would likely have succeeded in the German Physics Olympiad—a prototypical STEM competition—are missed by it. The study sample comprised 282 Olympiad participants and 1,103 non-participants from academic-track secondary schools, all assessed on 31 indicators spanning sociodemographic and educational background, cognitive abilities, self-related beliefs, motivational and affective variables, personality traits, vocational interests, and prior participation across other German STEM competitions. Three machine learning models (elastic net, random forest, gradient boosting) were trained on the participant sample to predict first-round success based on those indicators. The elastic net was found to perform best. Although individual-level predictive accuracy was modest, the model proved well-calibrated, supporting valid group-level inferences. Only a few variables predicted success: mathematics skills, physics- and engineering-related skills, and having skipped a grade were positive predictors, whereas conventional vocational interests were a negative predictor. When applied to the non-participants at a high-confidence threshold, the model identified 43 students (3.9%) as potentially successful. Crucially, 74% of these students had previously entered at least one selective STEM competition, most often the German Mathematics Olympiad, leaving only 11 students (≈ 1% of all non-participants) entirely unreached by the selective STEM competition system. These findings indicate that the majority of potentially successful non-participants are not absent from the selective STEM competition system but are engaged by it through other domains.

# INTRODUCTION

STEM (science, technology, engineering, and mathematics) competitions are well-established extracurricular enrichment activities for STEM-interested secondary school students worldwide (Beutel & Tetzlaff, 2007; Racherbäumer & Boltz, 2012). Heavily subsidized by governments (e.g., Stake & Mares, 2001) and reaching several million students worldwide each year, such competitions are designed to foster STEM potential during a developmental period in which adolescents form academic identities and make career-relevant choices (e.g., Steegh et al., 2019). For those who participate, prior research suggests that competitions can indeed be beneficial: former participants are more likely to pursue STEM careers, even when controlling for prior STEM interest (Miller et al., 2018), and retrospective studies consistently report that participants view their competition experience as formative for their professional development (Campbell & Walberg, 2011; Resch, 2013; Smith et al., 2021). For STEM competitions to have such positive consequences, students must participate. Students with the potential to succeed who do not participate may miss opportunities to further develop their STEM-related abilities and interests, and competitions may fail to reach part of the target group they are intended to identify and support.

Whether potentially successful non-participants exist is a question that requires understanding two aspects: what makes students successful in STEM competitions, and what leads them to participate in the first place. Research on STEM competitions has shown that domain-specific cognitive abilities, together with competence beliefs such as expectancy of success and self-efficacy, are central predictors of success in STEM competitions (e.g., Stang et al., 2014; Tschisgale, Steegh, Petersen, et al., 2024; Urhahne et al., 2012). Other research has also found a broader set of motivational, social, and contextual factors to influence participation. Students' participation is shaped by interest and self-efficacy, but also by teacher support, a conducive home environment, and gender-related identity processes, including stereotypic beliefs (e.g., Blankenburg et al., 2016; Campbell & Feng, 2010; Steegh et al., 2021b, for an overview, see Steegh et al., 2019). The two sets of relevant variables overlap only partly: the cognitive abilities that strongly explain success appear to play a smaller role in explaining participation, whereas many of the factors that drive participation seem to be largely absent from the set of variables that explain success. This divergence suggests that a student's likelihood of participating is only weakly informative about their likelihood of succeeding, and vice versa, though the two are not fully independent. It therefore raises the possibility that some students in the general school population possess characteristics associated with competition success but, lacking the characteristics that typically lead to participation, never enter the competition.

To examine this possibility, the present study uses a counterfactual predictive modelling approach that links competition participant and non-participant data. First, using participant data from the German Physics Olympiad on a set of variables (sociodemographic and educational background, STEM-related skills, self-related beliefs, motivational and affective variables, personality traits, and vocational interests), a supervised machine learning classifier is trained to predict competition success, defined as advancement beyond the first round of the Physics Olympiad. Second, the trained classifier is applied to non-participant data on the same variables to estimate each non-participant's counterfactual probability of first-round success in the competition. From these estimates, the subset of non-participants who would likely have been successful had they participated—the potentially successful non-participants—is identified. Third, prior competition experience of the identified potentially successful non-participants is examined to estimate how many of them have already been engaged by the broader selective STEM competition system through other pathways and how many have remained unreached.

# THEORETICAL BACKGROUND

## STEM Competitions and Their Landscape

STEM competitions are competitive informal learning activities that come in a broad range of formats with different foci. They commonly share the broader objective of supporting participating students in developing competence in a specific STEM domain and nurturing their motivation to possibly engage in STEM-related careers (Campbell et al., 2000; Petersen et al., 2017).

STEM competitions can be broadly categorized into two main formats: Project-centered competitions, such as science fairs and research project competitions, require participants to work individually or in teams on self-selected projects, thereby applying scientific methodologies to investigate a research problem of their own choosing (e.g., Jaworski, 2013; Marchak et al., 2026). These competitions particularly emphasize creativity, self-directed learning, and the ability to communicate scientific findings. Task-centered competitions, in contrast, such as the national and international Science Olympiads, take a different approach: participants work individually—often in exam-like settings—on demanding theoretical and often also experimental problems. These task-centered competitions often consist of multiple rounds, successively reducing the number of participants across rounds, ultimately identifying the most capable students. Although both formats aim to foster engagement in STEM, they place different demands on participants: Project-centered competitions emphasize scientific inquiry and communication skills, while task-centered competitions emphasize domain-specific problem-solving skills and knowledge.

Another way to categorize competitions is by the level of engagement they require from students, distinguishing between broad-participation and selective competitions. Broad-participation competitions are designed to reach as many students as possible: they are generally completed in a single brief session and typically administered in class, so that participation may reflect a school or class-level decision rather than a student's own choice. Selective competitions, by contrast, require students to actively decide to participate, often at a teacher's encouragement, and demand sustained self-directed engagement over weeks or months. Whereas participation in broad-participation competitions carries limited inferential value regarding individual motivation or domain commitment, prior participation in selective competitions may serve as a behavioral indicator of sustained interest and self-directed engagement in a domain.

Importantly, individual STEM competitions rarely operate in isolation. In many national contexts, they are instead embedded within a larger, coordinated ecosystem of task-centered as well as project-centered competitions that spans several domains and levels of engagement. By linking entry-level, broad-participation competitions with selective, multi-round competitions across adjacent domains, such an ecosystem can reach a larger share of STEM-interested students and offer multiple entry points through which individual potential may be recognized and supported. A student who does not enter one competition may therefore still be reached by another—a possibility that becomes analytically meaningful only when the landscape is treated as an interconnected system rather than as a set of isolated programs.

The German STEM student competition landscape presents one example of such an ecosystem. It is rather dense and institutionally coordinated: nationally advertised competitions are held to shared quality criteria, are formally recommended by the Standing Conference of the Ministers of Education and Cultural Affairs (KMK, 2009), and many receive public funding. Beyond these shared standards, the organizers also exchange experience and cooperate across competition boundaries. Within STEM alone, the landscape comprises a large number of competitions, ranging from broad entry-level formats to selective, multi-round Olympiads. Although the present study focuses on the German case, similarly coordinated competition systems exist in many other countries, so the systemic perspective adopted here is likely to extend beyond the German context.

## Research on STEM Competitions

### *Outcomes of Participation*

Research on STEM competitions has extensively investigated how participation influences the future careers of participating students. Studies by Resch (2013) and Smith et al. (2021) revealed that former participants believed that their competition experience had positively influenced their academic and career trajectories. Similarly, Miller et al. (2018) found that students who participated in STEM competitions were more likely to engage in a STEM-related career, even when controlling for prior STEM interest. Moreover, highly successful participants—those who performed best at a national level—were found far more likely to perform exceptionally well during their studies and career (Campbell, 1996; Campbell & Walberg, 2011), with longitudinal evidence pointing to elevated rates of scientific publications and patents (Campbell et al., 2000; Wai et al., 2010).

### *Predictors of Success*

A large body of research has also aimed to understand what variables are related to success in STEM competitions. Successful participants were consistently found to be highly interested in the competitions' respective domains (e.g., Forrester, 2010; Höffler et al., 2019; Lind, 2001). Campbell and Feng (2010), for example, found that less successful participants were often characterized by a lack of motivation. Self-concept of ability was identified as another feature distinguishing more and less successful participants (Campbell, 1996; Czerniak, 1996; Steegh et al., 2021b). Successful participants also often reported perceiving regular school classes as boring, suggesting they are under-challenged by regular schooling (Verna & Feng, 2002). Social influences mattered as well: successful participants generally came from families with conducive home atmospheres and reported strong parental and teacher support (Campbell & Feng, 2010; Lind & Friege, 2001; Steegh et al., 2021a). In terms of cognitive characteristics, general cognitive abilities (e.g., IQ) were not associated with success in advanced rounds of competition (Lind & Friege, 2001; Stang et al., 2014; Urhahne et al., 2012), whereas domain-specific cognitive abilities (such as domain-specific problem-solving abilities) were shown to play a more prominent role (Tschisgale, Steegh, Kubsch, et al., 2024; Tschisgale, Steegh, Petersen, et al., 2024).

Three studies are particularly relevant because they explicitly modeled competition success — operationalized as advancement at a specific competition round — as a function of multiple predictor variables: Urhahne et al. (2012) investigated advancement from the third to the fourth round of the German Chemistry Olympiad ($N = 52$) and identified previous participation as a significant predictor. Stang et al. (2014) examined advancement from the third to the fourth round of the German Biology and Chemistry Olympiads ($N = 87$), finding that boredom in school and expectancy of success predicted advancement. Both studies focused on later competition rounds, where participants had already passed several selection rounds, and relied on relatively small samples. Tschisgale et al. (2024) extended this line of research to earlier rounds of the German Physics Olympiad ($N = 136$). By including physics problem-solving ability alongside non-cognitive predictor variables, the study showed that the domain-specific cognitive ability was the primary predictor of success in the first and second round of the Physics Olympiad.

Although these studies were framed in terms of predicting success, their analytical orientation was primarily explanatory. Their focus was on identifying variables with statistically significant, non-zero associations with advancement in a specific round, rather than on evaluating whether these variables could jointly predict advancement with sufficient accuracy.

However, statistically significant predictors do not necessarily combine into a model with strong predictive performance. If the aim is to forecast which students are likely to succeed, a predictive modeling approach is required.

### *Predictors of Participation*

While research on success has focused on who succeeds among those who participate, a separate strand of research has examined what drives the decision to participate in STEM competitions in the first place. In their systematic review, Steegh et al. (2019) synthesized findings from 46 publications on mathematics and science competitions and found that the social environment—including parental, teacher, and peer influences—to be among the most frequently reported factors explaining participation. Blankenburg et al. (2016) found that students' willingness to participate was predicted by science interest, positive competition self-concept, and prior competition experience. Teacher support appears particularly important because teachers typically serve as the initial point of contact encouraging students to register (Abernathy & Vineyard, 2001; Tschisgale, 2024). Whether students ever learn of a selective competition therefore depends in part on how far their school engages with such competitions in the first place. Moreover, Höffler et al. (2017) found that participants of a science competition were more learning goal oriented, had less performance avoidance goals, and showed less work avoidance than comparable non-participants.

Gender-related factors add another dimension to the participation question. Steegh et al. (2019) documented substantial gender differences across STEM competitions, with female students particularly underrepresented in physics and chemistry, and suggested that stereotypes contribute to these disparities by shaping students' self-concepts and interests. In a later study, Steegh et al. (2021b) explicitly showed that implicit gender-science stereotypes played a role in the underrepresentation of girls in the Chemistry Olympiad, mediated through effects on self-concept and interest.

Together, these findings paint a picture in which participation in STEM competitions is not simply a function of motivational-affective variables but is substantially shaped by the social environment (parental, peer, and especially teacher influences) as well as by broader school-level factors (e.g., school-level awareness of competitions), self-related beliefs, and identity-related processes.

### *Mismatch Between Success and Participation Predictors*

The possibility of potentially successful non-participants becomes apparent when findings on predictors of participation and findings on predictors of success are considered together. Succeeding in a competition necessarily involves two steps: students must first enter, and then, given entry, perform well enough to advance (which is considered as success here). In the first step, motivational, social, and contextual factors — interest in the domain, self-related beliefs, teacher and parental encouragement, school-level awareness, and gender-related identity processes — take the primary role in shaping a student's decision to participate. In the second step, domain-specific cognitive skills, prior achievements, and competence beliefs are central in determining whether a participant advances.

Although entry and advancement are distinct steps, the variables that predict them are not fully independent; three mechanisms in particular link the predictors of entry to those of success. First, motivation shapes the amount of time students invest in learning and thereby contributes to domain-specific expertise development (Bransford et al., 2000). Accordingly, students with stronger domain-related motivation may also show stronger domain-specific cognitive skills; conversely, highly able competition participants are unlikely to show no domain interest or motivation at all (Tschisgale, Steegh, Kubsch, et al., 2024). Second, teachers act as central gatekeepers for entry into the competition (Abernathy & Vineyard, 2001; Tschisgale, 2024) and typically encourage students based on perceived domain-specific ability, so cognitive factors may also influence entry through teacher-mediated recruitment. Third, self-related beliefs such as domain-specific self-concept are themselves shaped, in part, by prior achievement and ability (Marsh & Martin, 2011), so the entry-relevant self-concept variables are not strictly independent of the success-relevant cognitive variables either. As a consequence, participants tend to differ from non-participants not only on the motivational, social, and contextual variables that drive entry but also, to some extent, on the cognitive variables relevant for success.

Because the relationship between participation-relevant and success-relevant variables is probabilistic rather than deterministic, some students may possess the domain-specific cognitive characteristics associated with competition success without necessarily entering competitions. Such students are unlikely to be entirely disconnected from the domain, because advanced domain-specific skills usually require some degree of prior interest, engagement, or sustained learning (Bransford et al., 2000). Yet domain-related engagement does not necessarily lead students to enter competitions. Some students may be interested in the domain but not in competing, may doubt whether they belong in a competition setting, or may simply not receive teacher encouragement or information about the opportunity. Such students — non-participants who could have succeeded had they participated — are referred to as potentially successful non-participants in this study.

Thus, when these potentially successful students do not participate, this may represent a missed opportunity to support their STEM-related development. As reviewed above, competition participation can provide formative experiences for academic and career development; students who could benefit from and succeed in such competitions but never enter may miss opportunities for challenge, recognition, and further support. Yet the magnitude and meaning of this missed opportunity depend on whether these students are absent only from a given competition or from the broader STEM competition ecosystem altogether.

In the first scenario, potentially successful non-participants in the focal competition (e.g., in the German Physics Olympiad) may already be engaged elsewhere in the selective STEM competition landscape, for example through another STEM competition. The broader ecosystem would then have reached these students, but not through the competition in question. In this scenario, non-participation in a specific competition would not necessarily indicate a failure of the broader potential-development ecosystem, but rather a domain-specific distribution of participation across available pathways. In the second scenario, potentially successful non-participants may have no prior experience with selective STEM competitions at all. They would then remain unreached by the competition ecosystem as a whole, pointing to a more fundamental gap in awareness, access, or recruitment.

This distinction motivates the present study, which examines not only to what extent such potentially successful non-participants exist, but also how they have already been reached through other STEM competition pathways or not.

**The Present Study**

Research has established what predicts success among STEM competition participants and what shapes participation among the broader student population. Yet the intersection of these two questions remains largely unexamined: whether some non-participating students nevertheless show characteristics associated with competition success. Addressing this question requires a counterfactual perspective, because these students' actual competition performance is unobserved. It is therefore necessary to estimate how likely non-participants would have been to succeed had they participated.

The present study addresses this question in the context of the German Physics Olympiad (for details, see Petersen & Wulff, 2017). The study used data from the *WinnerS* research project, which includes comprehensive measures of sociodemographic and educational background, STEM-related skills, self-related beliefs, motivational and affective characteristics, personality traits, vocational interests, and prior competition participation for both Physics Olympiad participants and non-participating students assessed with the same instruments. Using supervised machine learning, predictive models were trained on participant data to predict first-round success, and the best-performing model was then applied to the non-participant sample. This allowed to identify potentially successful non-participants and examine whether they had already been reached through other selective STEM competition pathways beyond the Physics Olympiad.

Specifically, the following research questions (RQs) were asked:

**RQ1:** To what extent can first-round success in the German Physics Olympiad be predicted from participants' sociodemographic and educational background, STEM-related skills, self-related beliefs, motivational-affective variables, personality traits, and vocational interests, and which characteristics contribute most to the prediction?

**RQ2:** How many potentially successful non-participants can be identified, that is, students who would likely have succeeded had they participated?

**RQ3:** How many of the identified potentially successful non-participants have previously entered other selective STEM competitions, and what share remains completely unreached?

# METHODS

## Study Context and Data Collection

The present study focuses on the German Physics Olympiad—a selective task-centered student competition (for details, see Petersen & Wulff, 2017). This annual competition for secondary school students consists of four selection rounds involving predominantly theoretical and, to a lesser extent, experimental physics problems. In the first round, approximately 900 students voluntarily submit solutions to problems worked on individually at home over about five months. Students scoring above a predefined threshold advance to the second round; typically, 40–60% do so. The second round is a theoretical examination administered at participants' schools, while the third and fourth rounds each involve a one-week in-person stay at a research institute comprising theoretical and experimental examinations alongside seminars, excursions, and talks. The top five students in the fourth round represent Germany at the annual International Physics Olympiad.

Despite cross-national differences in scale and implementation, the German Physics Olympiad captures key structural features that characterize selective task-centered STEM competitions internationally: multi-round selection procedures, a shift from decentralized to centralized rounds, and a growing emphasis on experimental problem solving in later rounds (not for mathematics though). Moreover, because STEM competitions—irrespective of their precise domain—rely on shared scientific practices, the German Physics Olympiad can serve as a plausible prototype for examining questions relevant to STEM competitions more broadly.

The data used in this study were drawn from a larger research project (called *WinnerS*) that investigated predictors of success and failure in major German science competitions (including the Physics Olympiad). Students participating in one of the relevant competitions in 2018 could voluntarily take part in a concurrent online questionnaire that assessed students' sociodemographic and educational background, STEM-related skills, self-related beliefs, motivational and affective variables, certain personality traits, vocational interests, as well as prior participation in STEM competitions. The same data were collected from non-participating students who served as a control group, matched in terms of school type and grade level. Data collection for the control group followed a traditional pen-and-paper format.

## Sample

The study sample comprises two groups: The *participant* group consists of 282 students who participated in the German Physics Olympiad and in the concurrent online questionnaire, whereas the *non-participant* group consists of 1,103 students from academic-track secondary schools (*Gymnasium*) in Germany. Restricting the non-participant group to academic-track schools was justified, given that 96.5% of Physics Olympiad participants in our sample with available information on school type attended an academic-track school.

The two groups were comparable in grade level: the mean grade level was 11.2 among Olympiad participants (*SD* = 1.0) and 11.0 among non-participants (*SD* = 0.7). The groups differed in gender composition: the non-participant group was approximately gender-balanced, (49% female students), whereas the Olympiad participant group included a lower proportion of female students (33% female students).

Of the 282 Olympiad participants in our study, 132 (46.8%) advanced beyond the first round and 150 (53.2%) did not. In the overall Physics Olympiad population for that year, 406 of 931 first-round participants (43.6%) advanced to the second round. Thus, the participant sample in the present study appears to have been slightly positively selected with respect to first-round success.

**Instruments**

*Sociodemographic and educational background* was assessed through self-report items covering grade level, gender, whether students had been formally identified as intellectually gifted, whether they had ever skipped a grade, and whether a non-German language is spoken at home. Socioeconomic status was measured via the number of books available at home, a widely used proxy adapted from PISA 2003 (OECD, 2007).

*STEM-related skills* were captured using two single-choice skill tests: a physics and engineering-related skills test with 15 items drawn from Heller and Perleth (2007), and a self-developed 24-items mathematics skill test. The latter covered topics up to secondary school mathematics (e.g., algebraic manipulation, basic geometry, and integral calculus) as well as topics typically encountered in early undergraduate studies, such as partial derivatives, complex numbers, vector calculus, Taylor series, and ordinary differential equations.

*Self-related beliefs* encompassed five constructs. Physics self-concept and mathematics self-concept were separately assessed with six items each adapted from PISA 2003 (OECD, 2007). Academic and career success expectations were measured with four items selected and adapted from Lykkegaard and Ulriksen (2016) and Eccles and Wigfield (1995) and translated into German. Entity theory of intelligence was assessed using the scale by Froehlich et al. (2016). Gender stereotypes in science were measured with four items adapted from the Fennema–Sherman Mathematics Attitudes Scales (Fennema & Sherman, 1976) and translated into German.

*Motivational and affective variables* comprised six constructs. Subject interest and topic interest in physics were each measured with four items from Daniels (2008), capturing students' affective orientation toward physics as a discipline and toward specific physics content, respectively. Physics career motivation was assessed with three items adapted from Urhahne et al. (2012), targeting long-term career aspirations. Goal orientations were measured using the four subscales of the SELLMO (Spinath et al., 2002) — mastery goals, performance-approach goals, performance-avoidance goals, and work-avoidance goals — with four items retained per subscale and adapted to science classes. Boredom in physics classes was assessed with four items originally developed for PISA 2003 (OECD, 2007) and adapted to physics.

*Personality traits* were represented by two constructs: grit, assessed with eight items selected from the original 12-item scale (Duckworth et al., 2007) and translated into German; and conscientiousness, measured using all three items from the conscientiousness subscale of the Big Five Inventory–SOEP (Schupp & Gerlitz, 2008).

*Vocational interests* were assessed using the RIASEC+N framework (Dierks et al., 2014), which adapts Holland's classic typology to school settings and extends it with a Networking dimension, yielding seven subscales with four items each. Political engagement was measured using the full five-item scale by Otto and Bacherle (2011).

*Success in the competition* (more precisely, success in the first round of the Physics Olympiad) was operationalized as a binary outcome: advancement to the second round or beyond (coded as 1) versus non-advancement (coded as 0). Advancement decisions were based on participants' scores on the submitted solutions for the first round. A predetermined threshold was applied, with students scoring above the threshold advancing and those scoring below it not advancing.

*Prior competition experience* was assessed via a single item asking students to indicate which of 12 named German STEM competitions they had previously participated in (multiple responses permitted; an open "other" category was included). The listed competitions span two qualitatively distinct formats: broad-participation competitions in which typically entire classes or schools participate, sometimes on a mandatory basis (e.g., Mathematical Kangaroo competition), and selective competitions in which individual students — often encouraged by a teacher — decide to participate on their own initiative (e.g., Jugend forscht, Mathematics Olympiad, Physics Olympiad; for the full list, see Supplemental Material Part A).

## Analytical Approach

All analyses were implemented in Python (version 3.10).

### *Predicting Success and Identifying Important Predictors (RQ1)*

*Candidate supervised machine learning classifiers.* Three supervised machine learning classifiers were trained to predict success in the first round of the German Physics Olympiad based on a large set of available predictor variables, referred to as features in machine learning terminology. First, an elastic net logistic regression classifier was included as a simple linear model (Zou & Hastie, 2005). Second, a random forest classifier was included — a tree-based ensemble aggregating predictions across multiple decorrelated decision trees (Breiman, 2001a). Unlike elastic nets, random forests can capture nonlinear effects of predictors and higher-order interactions without manual specification, making them well suited for a broad predictor set with unknown functional relationships. Third, a gradient boosting machine was included as a strong and flexible nonlinear alternative that tends to achieve higher predictive accuracy than random forests in practice, albeit at the cost of greater tuning complexity (Natekin & Knoll, 2013).

*Model training, evaluation, and selection.* A nested cross-validation approach was used to obtain unbiased performance estimates of the three models while simultaneously tuning hyperparameters (Stone, 1974). The outer loop used a 10-times repeated stratified 5-fold cross-validation (50 evaluation rounds in total), applied identically to all three candidate machine learning models. Within each outer training fold, an inner 5-fold stratified cross-validation selected the best hyperparameter combination for each model via grid search[1]. The Brier score was selected as the optimization criterion as it rewards accurate probability estimates rather than merely correct classifications (Brier, 1950). This is particularly relevant for this study given that predicted probabilities, rather than hard class assignments (i.e. successful vs. not successful), drive the subsequent identification of potentially successful non-participants. Missing values were handled through median imputation, replacing missing entries with the feature median computed exclusively on the training data, thereby preventing data leakage from test folds (for the extent of missingness across all variables, see the Supplemental Material Part C). Performance across the 50 outer folds was summarized using mean Brier score, AUC (area under curve of respective receiver operating characteristic curves), and Cohen's $\kappa$ (evaluated at a decision threshold of $p = .50$).

[1] Hyperparameter grids for all three models are provided in the Supplemental Material (Part B, Tables S2–S4).

The candidate machine learning classifier with the lowest mean outer-fold Brier score was selected as the best-performing model. It was then retrained on the full participant sample using a fresh 5-fold stratified cross-validation grid search (again optimizing the Brier score) to determine the final hyperparameters. This final model was used for all subsequent analyses.

*Calibration analysis of the final model.* A calibration analysis then examined whether the model's predicted probabilities could be interpreted as expected event rates. More specifically, participants assigned a probability $p$ of advancing by the model should actually advance at a frequency close to $p$ in the observed data. Calibration is particularly important for RQ2 because the identification of potentially successful non-participants does not rest on classifying individual students as either successful or unsuccessful but on interpreting predicted probabilities as expected success rates at the group level. Only under calibration can a predicted probability be read as the rate at which comparable students would actually advance.

Calibration was assessed using the per-participant cross-validation predictions, which provide out-of-sample estimates for each participant. More specifically, a calibration curve was inspected, which plots observed advancement rates against predicted probabilities. Calibration intercept and slope were further computed by regressing observed outcomes on logit-transformed predictions in a logistic regression (Cox, 1958): under perfect calibration, the intercept would equal zero and the slope would equal one. An intercept above zero indicates that observed advancement rates are systematically higher than predicted, while an intercept below zero indicates the opposite. A slope below one indicates that predictions are too extreme—high probabilities overshoot and low probabilities undershoot the corresponding observed rates—while a slope above one indicates the opposite, with predictions compressed toward the base rate. Lastly, as a global test, Spiegelhalter's $z$-statistic was computed to assess overall miscalibration (Spiegelhalter, 1986).

*Features driving predictions of the final model.* To probe the model's internal decision logic, SHAP (SHapley Additive exPlanations) values were computed for the final model on the participant sample (Lundberg & Lee, 2017; Molnar, 2023). SHAP values decompose each prediction into additive feature contributions at the level of individual observations, with the baseline taken as the average predicted success probability in the first round across participants. SHAP values were computed in the model's predicted-probability output space rather than log-odds, so individual contributions can be read directly as percentage-point shifts in predicted success probability. Examining whether the features driving predictions align with prior literature on competition success additionally serves as a post-hoc plausibility check of the predictive model.

### *Identifying Potentially Successful Non-Participants (RQ2)*

*Counterfactual prediction of competition success*. The final model was applied to the non-participant data. Missing values were imputed using the median values derived from the participant training data to prevent data leakage. Each non-participant was thereby assigned a predicted probability of advancing beyond the first competition round. Transferring the model in this way assumes that the relationship between the measured features and first-round success among participants also holds among non-participants; this assumption cannot be verified directly, since non-participants have no observed competition outcome. It is likely to be most strained for any non-participants whose measured characteristics fall outside the range well represented in the participant sample, where the predicted probabilities would rest on extrapolation rather than interpolation.

*Decision threshold selection.* Identifying potentially successful non-participants requires selecting a decision threshold $p$ above which a student is considered as potentially successful. Precision was prioritized over recall since false positives would erode the construct validity of the potentially successful non-participant group more severely than false negatives. Hence, the threshold $p$ was selected empirically using a precision-targeting approach, targeting a precision proxy of $\approx 80\%$. Across thresholds from $p = .001$ to $p = .999$ (step = .001), an empirical precision proxy was computed, i.e. the actual success rate among participants whose predicted probability exceeded that value. Because this proxy is derived from participants, it approximates the precision expected among potentially successful non-participants only if model calibration transfers across the two groups. The identified group should therefore be interpreted as a pool of potentially successful non-participants rather than as a reliable classification of specific individuals.

### *Investigating Prior Competition Experience (RQ3)*

*Prior competition experience and identification of unreached potentials.* Prior competition participation was examined to determine whether potentially successful non-participants had already been reached by the selective STEM competition system through other competition pathways. Prior participation among the remaining non-participants who were not classified as potentially successful served as a descriptive baseline for typical STEM competition experiences among ordinary students, allowing to interpret the participation experiences of potentially successful non-participants against this baseline.

Prior participation was assessed separately for selected major broad-participation and selective STEM competitions in Germany. Broad-participation competitions (such as the Mathematical Kangaroo and the Bebras Competition) were included but not used to define STEM competition system reach, because they are often administered at class or school level and therefore do not necessarily reflect individual initiative or selective engagement. System reach was instead operationalized as any prior participation in the selected selective competitions. Potentially successful non-participants with no recorded prior participation in any of these selective STEM competitions were defined as unreached potentials. Thus, this group represents students whose predicted potential for success in the Physics Olympiad had not previously been engaged through the considered selective STEM-competition pathways. A full list and further information of all considered STEM competitions in RQ3 can be found in the Supplemental Material (Part A).

# RESULTS

## Scale Characteristics and Descriptive Group Comparisons

Table 1 provides scale characteristics and descriptive information for all study variables, including the number of items and internal consistencies measured by Cronbach's α, as well as descriptive statistics for Physics Olympiad participants and non-participants. Internal consistencies of all variables proved acceptable (Hair et al., 2019). Given the large sample size, almost all group comparisons reached statistical significance; therefore, any comparison of both groups should primarily be based on effect sizes. The largest effect sizes, with $|d| \geq 1.0$, favored Physics Olympiad participants and were observed for STEM-related skills, physics-specific motivational-affective variables (subject interest in physics, topic interest in physics, and physics career motivation), as well as physics and mathematics self-concept.

**Table 1.** Scale characteristics and descriptive information for all study variables. Scale characteristics include the response scale, number of items, and internal consistency estimates assessed using Cronbach's α. Descriptive information compare Physics Olympiad participants and non-participants across all study variables, including group means, standard deviations, effect sizes reported as Cohen's *d*, and exploratory *p*-values.

| Construct | Scale | No. of Items | Cronbach's *α* | *M*(*SD*) of Participants | *M*(*SD*) of Non-Participants | Cohen's *d* | *p* |
|---|---|---|---|---|---|---|---|
| *Sociodemographic and educational background* | | | | | | | |
| Grade level | 8,…,13 | 1 | — | 11.19 (0.98) | 11.00 (0.73) | +0.24 | .002 |
| Gender (1 = female) | 0, 1 | 1 | — | 0.33 (0.47) | 0.49 (0.50) | -0.33 | <.001 |
| Non-German home language (1 = yes) | 0, 1 | 1 | — | 0.08 (0.27) | 0.11 (0.32) | -0.10 | .147 |
| Socioeconomic status (based on owned books) | 1,…, 6 | 1 | — | 4.73 (0.67) | 4.30 (0.95) | +0.48 | <.001 |
| Identified giftedness (1 = yes) | 0, 1 | 1 | — | 0.23 (0.42) | 0.12 (0.32) | +0.32 | <.001 |
| Skipped a grade (1 = yes) | 0, 1 | 1 | — | 0.06 (0.24) | 0.04 (0.20) | +0.10 | .180 |
| *STEM-related skills* | | | | | | | |
| Physics and engineering-related skills | 0,…,15 | 15 | 0.67 | 0.57 (0.20) | 0.35 (0.18) | +1.15 | <.001 |
| Mathematics skills | 0,…,24 | 24 | 0.70 | 0.37 (0.22) | 0.21 (0.14) | +1.03 | <.001 |
| *Self-related beliefs* | | | | | | | |
| Physics self-concept | 1–4 | 6 | 0.95 | 3.50 (0.51) | 2.50 (0.86) | +1.25 | <.001 |
| Mathematics self-concept | 1–4 | 6 | 0.95 | 3.60 (0.51) | 2.87 (0.78) | +1.00 | <.001 |
| Academic and career success expectations | 1–4 | 4 | 0.81 | 3.00 (0.54) | 2.50 (0.64) | +0.80 | <.001 |
| Entity theory of intelligence | 1–4 | 3 | 0.87 | 2.26 (0.85) | 2.26 (0.74) | 0.00 | .958 |
| Gender stereotypes | 1–4 | 4 | 0.77 | 1.49 (0.56) | 1.35 (0.53) | +0.27 | <.001 |
| *Motivational and affective variables* | | | | | | | |
| Subject interest in physics | 1–4 | 4 | 0.93 | 3.52 (0.50) | 2.34 (1.00) | +1.28 | <.001 |
| Topic interest in physics | 1–4 | 4 | 0.94 | 3.41 (0.59) | 2.19 (1.00) | +1.31 | <.001 |
| Physics career motivation | 1–4 | 3 | 0.96 | 3.20 (0.83) | 1.94 (1.02) | +1.28 | <.001 |
| Goal Orientation | | | | | | | |
| Mastery goals | 1–4 | 4 | 0.80 | 3.71 (0.41) | 3.42 (0.55) | +0.56 | <.001 |
| Performance-approach goals | 1–4 | 4 | 0.78 | 2.49 (0.74) | 2.45 (0.71) | +0.05 | .458 |
| Performance-avoidance goals | 1–4 | 4 | 0.83 | 1.68 (0.65) | 1.84 (0.70) | -0.22 | <.001 |
| Work-avoidance goals | 1–4 | 4 | 0.80 | 1.79 (0.62) | 2.10 (0.74) | -0.44 | <.001 |
| Boredom in physics classes | 1–4 | 4 | 0.92 | 1.63 (0.58) | 2.31 (0.99) | -0.74 | <.001 |
| *Personality traits* | | | | | | | |
| Grit | 1–4 | 8 | 0.76 | 2.82 (0.48) | 2.72 (0.50) | +0.21 | .002 |
| Conscientiousness | 1–4 | 3 | 0.60 | 3.04 (0.51) | 2.91 (0.60) | +0.23 | <.001 |
| *Vocational interests* | | | | | | | |
| RIASEC+N | | | | | | | |
| Realistic | 1–4 | 4 | 0.72 | 2.45 (0.60) | 2.43 (0.67) | +0.03 | .673 |
| Investigative | 1–4 | 4 | 0.76 | 3.14 (0.55) | 2.56 (0.70) | +0.85 | <.001 |
| Artistic | 1–4 | 4 | 0.61 | 2.45 (0.59) | 2.14 (0.64) | +0.50 | <.001 |
| Social | 1–4 | 4 | 0.70 | 3.05 (0.58) | 2.73 (0.69) | +0.48 | <.001 |
| Enterprising | 1–4 | 4 | 0.61 | 2.60 (0.58) | 2.24 (0.66) | +0.55 | <.001 |
| Conventional | 1–4 | 4 | 0.61 | 2.05 (0.61) | 2.06 (0.65) | -0.03 | .665 |
| Networking | 1–4 | 4 | 0.83 | 3.27 (0.56) | 2.60 (0.79) | +0.90 | <.001 |
| Political engagement | 1–4 | 5 | 0.93 | 2.70 (0.87) | 2.52 (0.88) | +0.20 | .004 |

*Note.* For the two STEM-related skills tests (i.e. physics and engineering-related skills, mathematics skills), α equals KR-20 and, given heterogeneous item difficulty, is a lower bound on the true reliability. *p*-values were computed from Welch's *t*-test, except for the dichotomous variables where Fisher's exact test was used.

## Predicting Success and Identifying Important Predictors (RQ1)

*Model comparison and final model training.* Three candidate supervised machine learning classifiers (elastic net logistic regression, random forest, and gradient boosting) were trained and compared via 10-times repeated stratified 5-fold nested cross-validation (50 outer evaluation folds in total). Table 2 presents mean performance across the 50 outer folds.

Given the near-equivalent performance of all three models, the elastic net was selected as the final model due to its greater parsimony as a simple linear classifier. Its hyperparameters were then tuned on the full participant sample ($N$ = 282) via 5-fold cross-validation optimizing the Brier score. The grid search selected $C$ = 0.10 with a pure LASSO penalty (l1_ratio = 1.0), and the elastic net was refit once on all 282 participants using these values. The LASSO penalty drives the coefficients of uninformative predictors exactly to zero, performing automatic feature selection: of the 31 available predictors, 26 were shrunk to zero, leaving 5 with non-zero coefficients in the final model[2].

[2] Because the L1 penalty performs feature selection, the retained predictor set can vary across resamples. Across the 50 outer cross-validation folds, mathematics skills and physics and engineering-related skills were retained in 100% of folds, RIASEC-Conventional in 82%, skipped-a-grade in 80%, and subject interest in physics in 42%,

**Table 2.** Cross-validated performance of the three candidate supervised machine learning classifiers.

| Machine learning classifier | Brier (BSS) | AUC | Cohen's $\kappa$ |
|---|---|---|---|
| Elastic net | 0.237 (0.048) | 0.605 | 0.177 |
| Random forest | 0.239 (0.040) | 0.605 | 0.172 |
| Gradient boosting | 0.244 (0.020) | 0.610 | 0.171 |

*Note.* Metrics are means across the 50 outer evaluation folds. Brier score: lower is better; baseline Brier (constant base-rate prediction) = 0.249. Brier Skill Score (BSS) = 1 − (model Brier / baseline Brier); higher is better; 0 = no improvement over baseline. AUC and Cohen's $\kappa$: higher is better; chance baselines are 0.50 and 0.00, respectively. Cohen's $\kappa$ was evaluated at a decision threshold of $p = .50$.

*Calibration analysis of the final model.* Figure 1 displays the calibration curve of the final elastic net model based on participant cross-validation predictions. The calibration curve deviates from perfect calibration in a non-linear pattern: It lies above the diagonal at the lower and upper ends of the predicted probability range, indicating that observed success rates exceeded predicted probabilities in these regions, whereas it falls below the diagonal in the middle range, indicating that observed success rates were lower than predicted probabilities.

Calibration intercept and slope were additionally estimated from a logistic calibration model regressing the observed outcomes on the predicted log-odds. The calibration intercept was indistinguishable from zero ($\beta_0 = 0.03$, $SE = 0.13$), indicating no overall offset between predictions and outcomes. The calibration slope was $\beta_1 = 1.22$ ($SE = 0.33$), consistent with mild compression of predicted probabilities toward the base rate — a common consequence of L1-type regularization (Pavlou et al., 2024). Spiegelhalter's $z$-statistic, which summarizes overall miscalibration in a single test, was non-significant ($z = –0.63$, $p = .53$). Quantifying the upper-tail pattern from Figure 1, participants exceeding the precision-relevant thresholds advanced at higher rates than their mean predictions, e.g., 79% vs. 67% above $p = .58$, and 83% vs. 69% above $p = .60$. In other words, participants identified by the model as likely to advance did so even more often than the model itself predicted.

whereas every other predictor was retained in 30% of folds or fewer. Full details are provided in the Supplemental Material (Part D).

**Fig. 1.** Calibration curve for the selected elastic net model predicting participant outcomes.

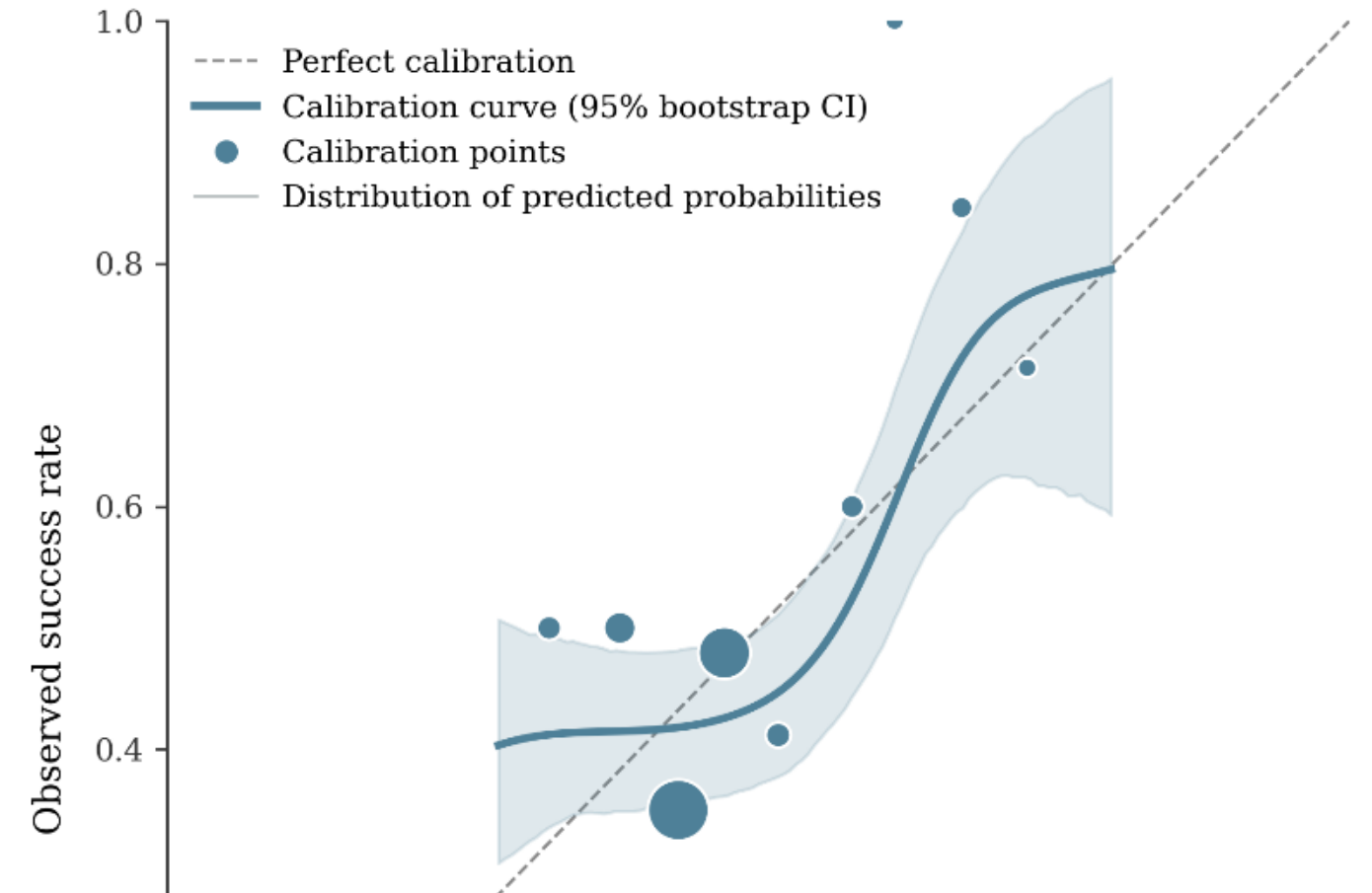

*Note.* Predicted probabilities of success are per-participant predictions of advancing beyond the first competition round, averaged across cross-validation repeats. The *y*-axis shows the corresponding observed proportions of advancement. The dashed line marks perfect calibration. The calibration curve is a kernel-smoothed estimate (Nadaraya–Watson, Gaussian kernel, bandwidth = 0.075) of observed success rate as a function of predicted probability of success, with a 95% bootstrap confidence band. Calibration points show observed success rates plotted against the mean predicted success probability within equal-width bins (0.05); marker size reflects the number of participants per bin. Gray spikes along the *x*-axis show the distribution of predicted probabilities of success.

*Features driving the prediction.* SHAP values were computed for the final elastic-net model applied to the participant data. Of the 31 input features, the elastic-net regularization retained five features with non-zero coefficients; the L1 (LASSO) component of the penalty shrank the remaining 26 features' coefficients to zero, essentially removing them from the model. Figure 2 shows the distribution of SHAP values across participants for the five retained features.

The SHAP analysis revealed mathematics skills as the dominant predictor, accounting for by far the largest spread of SHAP values, ranging from approximately −0.18 to +0.28 in predicted-probability units, with a clean monotonic pattern: high mathematics scores shifted predicted success substantially upward, while low scores shifted it downward by a comparable amount. The second-ranked feature, RIASEC-Conventional, showed a negative direction, with higher conventional vocational interest associated with decreased predicted probability of success.

Physics and engineering-related skills showed a similar pattern as mathematics skills; however, values ranged only between approximately −0.11 and +0.10. The dichotomous predictor skipped-a-grade was driven by a small number of participants who skipped a grade (visible as the isolated red dots at SHAP ≈ +0.05 to +0.07), and given the rarity of grade skipping (about 6% only), this contribution rests on few observations. The last feature not driven to zero by the elastic net is subject interest in physics, which, as Figure 2 conveys, has essentially no practical effect on the probability of success.

**Fig. 2.** SHAP feature contributions to predicted success probability among participants.

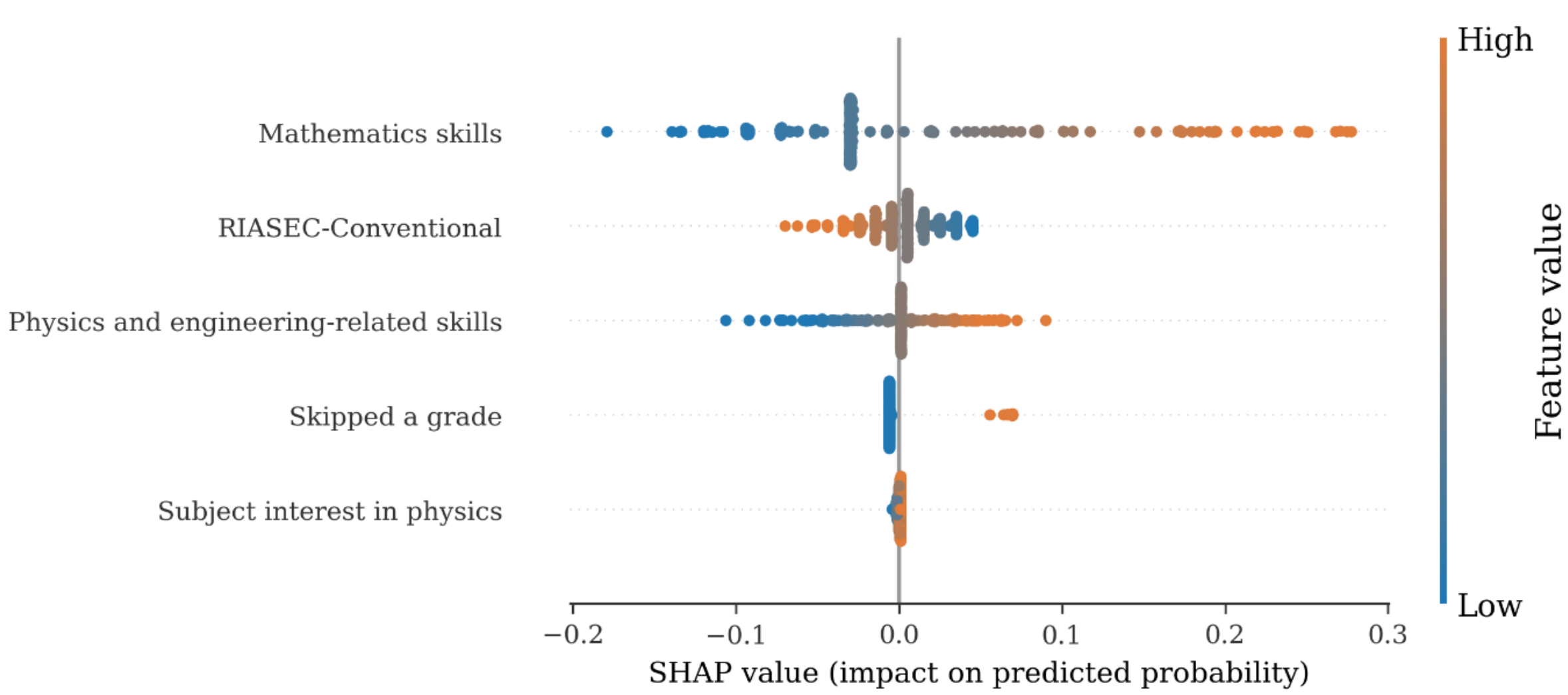

*Note.* Each row corresponds to one retained feature of the elastic net model, and each dot represents one participant. The dot's horizontal position shows the feature's additive contribution to the participant's predicted probability of advancing beyond the first competition round, relative to the model's average prediction. Positive SHAP values increase the predicted probability, whereas negative values decrease it. Color encodes the standardized feature value (orange = high, blue = low). Features are ordered by mean absolute SHAP value.

### Identifying Potentially Successful Non-Participants (RQ2)

*Distribution of predicted success probabilities.* The final elastic net model was applied to the 1,103 non-participants to estimate, for each, the probability of first-round success. Missing values were imputed using feature medians from the participant training data, mirroring the preprocessing applied during model training. Figure 3 shows the resulting distribution of predicted success probabilities for non-participants alongside the cross-validated predictions for participants. The two distributions differ in location: the non-participant distribution is shifted toward lower predicted success probabilities than the participant distribution. The mean predicted success probability among non-participants was 0.359, compared with 0.468 for participants (cross-validated predictions); the corresponding actually observed first-round success rate among participants was .468.

Participants in the upper region of the predicted-probability distribution can be regarded as those for whom the model predicts a high probability of first-round success. Identifying potentially successful non-participants therefore requires the choice of a decision threshold *p* above which a participant is considered likely to succeed, and applying this same threshold to the non-participant distribution to identify potentially successful non-participants (whose predicted probability of first-round success exceeds *p*).

**Fig. 3.** Distribution of predicted first-round success probabilities for non-participants and participants.

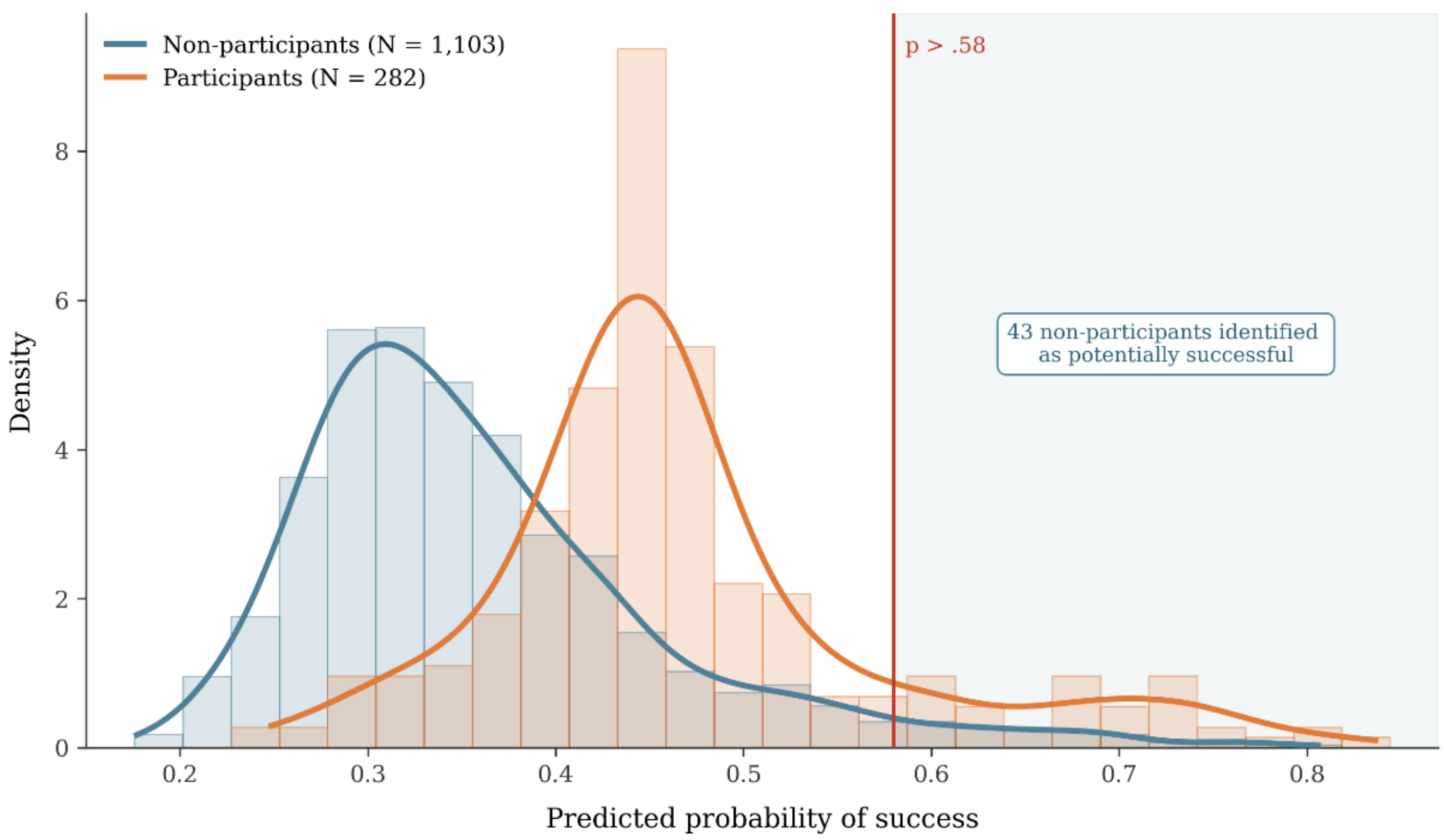

*Note.* Density curves are Gaussian kernel density estimates with bandwidth selected via Scott's rule. For participants, predicted probabilities are the cross-validated out-of-fold predictions from the nested cross-validation procedure (averaged across folds); for non-participants, they are predictions from the final model trained on the full participant sample. The vertical red line marks the threshold $p = .58$ used as the primary cut-off for identifying potentially successful non-participants.

*Threshold selection.* The precision-targeting procedure (≈ 80% target) selected a threshold of p = .58, at which the precision proxy reached 78.9% (see Figure 4). At this threshold, 43 non-participants (3.9% of the sample) were identified as potentially successful (see Figure 3).

**Fig. 4.** Selection of the decision threshold for identifying potentially successful non-participants.

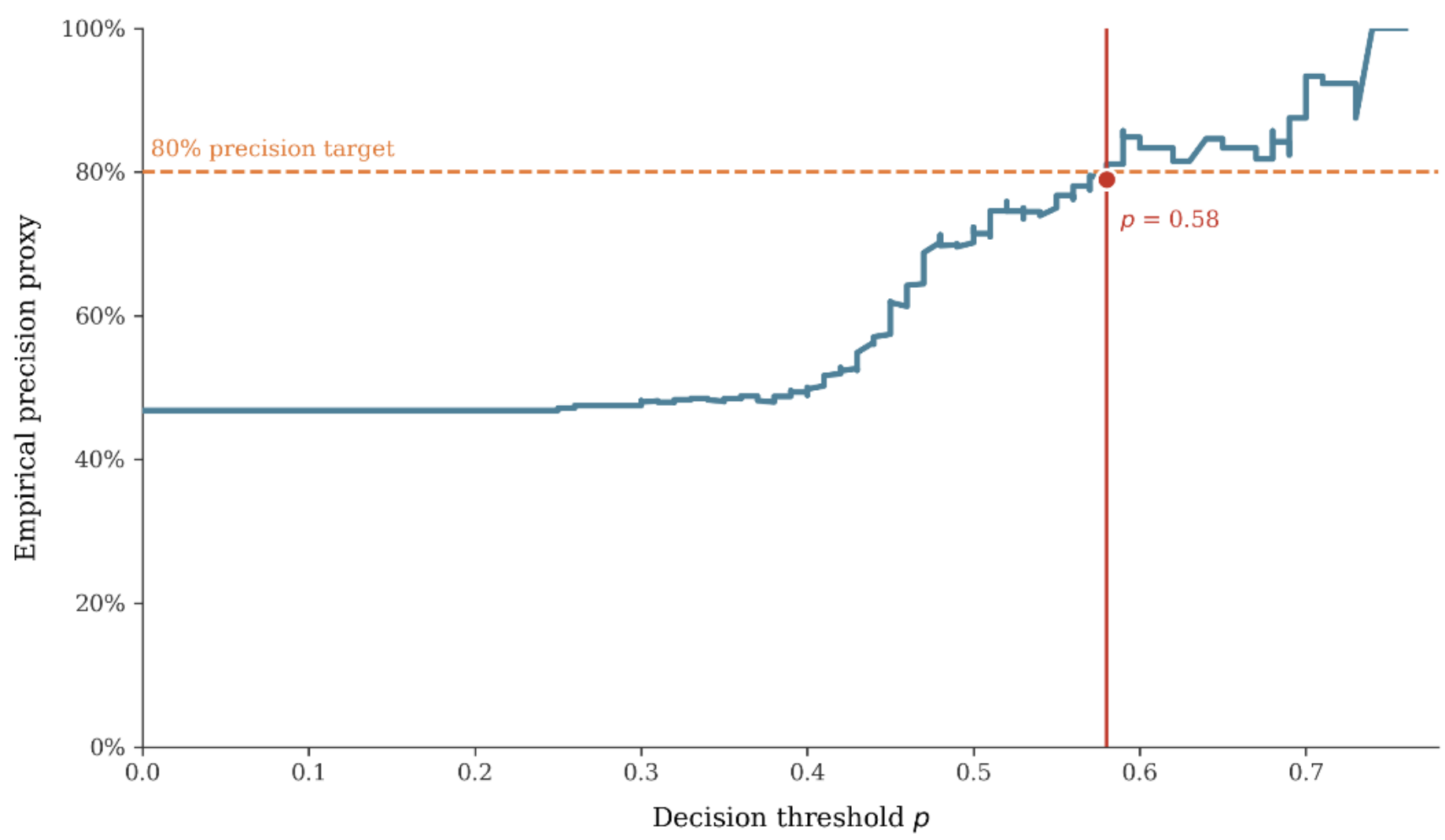

*Note.* The blue line shows the precision proxy: the observed first-round success rate among participants with predicted probability above the decision threshold $p$. The horizontal dashed line marks the 80% precision target. The vertical red line marks the selected threshold $p = .58$, at which the precision proxy is close to the 80% target (78.9%, $n = 43$).

## Investigating Prior Competition Experience (RQ3)

To address whether potentially successful non-participants were genuinely outside the selective STEM competition system or had already been engaged through other competition pathways, prior STEM competition participation of potentially successful non-participants was examined and benchmarked against the remaining non-participants who were not identified as potentially successful.

Fig 5 displays prior participation rates for both groups across all listed STEM competitions. Across nearly all selective competitions, potentially successful non-participants showed higher prior participation rates than the baseline group. The difference was largest for the focal competition itself: 30.2% of the potentially successful non-participant group had previously entered the German Physics Olympiad, compared with 3.7% of other non-participants. The German Mathematics Olympiad was the most common prior entry in both groups (55.8% vs. 38.2%). Differences in the same direction appeared for most remaining selective competitions (e.g., German National Computer Science Competition: 16.3% vs. 4.0%; German Junior Science Olympiad: 16.3% vs. 3.3%; German National Mathematics Competition: 14.0% vs. 2.9%). For two competitions the rates ran in the opposite direction, with small underlying counts: the German Biology Olympiad (2.3% vs. 4.6%) and the German National Environmental Competition (0.0% vs. 0.8%).

Aggregating across all listed selective competitions, 32 of the 43 potentially successful non-participants (74.4%) had at least one prior selective competition entry, compared with 47.6% of other non-participants. Conversely, 25.6% (11 of 43) of the potentially successful non-participant group had no such prior selective STEM competition experience, compared with 52.4% of other non-participants. These 11 students are referred to as *unreached potentials*. Prior participation in broad-participation competitions was high in both groups and higher among potentially successful non-participants (Mathematical Kangaroo: 86.0% vs. 71.7%; Bebras Competition: 76.7% vs. 37.6%).

**Fig. 5.** Prior STEM competition participation of potentially successful non-participants and other non-participants.

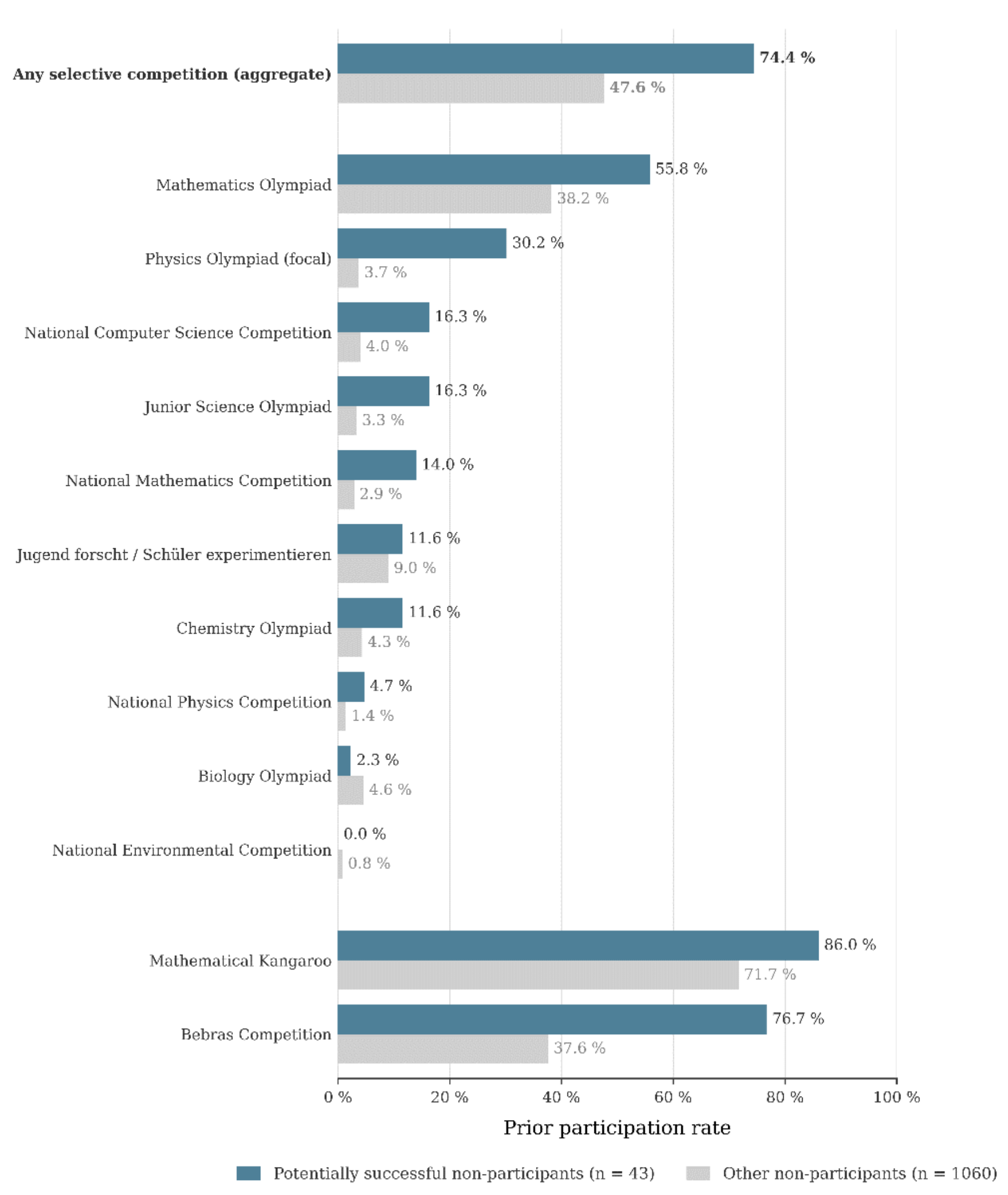

*Note.* Bars show the proportion within each group that reported prior participation in each listed STEM competition. Dark bars represent potentially successful non-participants ($n = 43$); light bars represent other non-participants ($n = 1{,}060$). The aggregate "Any selective competition" refers only to the listed selective STEM competitions and excludes the two broad-participation competitions (Mathematical Kangaroo, Bebras Competition), which are shown separately at the bottom. A full list of and further details on the considered STEM competitions are provided in the Supplemental Material (Part A).

# DISCUSSION

## Predicting Success and Identifying Important Predictors (RQ1)

RQ1 asked to what extent first-round success in the German Physics Olympiad can be predicted from a range of variables, and which variables contribute most to that prediction. To address RQ1, three supervised machine learning models were trained and compared. All trained models performed only modestly in predicting individual success in the first round of the Physics Olympiad, reflecting the well-discussed limits of predicting more distal social and educational outcomes (e.g., Narayanan & Kapoor, 2025). Provided the comparable performance of all three models, the most parsimonious model was retained for interpretation: an elastic net logistic regression model that effectively converged to a LASSO logistic regression. By shrinking the coefficients of irrelevant predictors to zero during training, the LASSO retained only five non-zero predictors, four of which differed meaningfully from zero. SHAP value analysis further indicated that mathematical skills were the most influential predictor of first-round success, followed by RIASEC-Conventional, physics and engineering-related skills, and whether participants had skipped a grade. Although the model's capability to predict first-round success for individual students was modest, the retained model was still informative in two respects: It showed that the predictive signal was concentrated in a small set of predictors, and it produced probability estimates that, in the range relevant to this study, are calibrated well enough to support group-level inferences.

The strong predictive contribution of mathematics skills is open to two interpretations. First, it may reflect the mathematical competence the competition directly demands: the Physics Olympiad relies heavily on mathematics (Treiber et al., 2023), and quantitative abilities have been identified as an important success predictor in the Physics Olympiad before (Tschisgale, Steegh, Petersen, et al., 2024). Second, it may partly reflect prior engagement with STEM content beyond the school curriculum. The instrument used here covered material typically encountered only in early undergraduate studies (e.g., partial derivatives, complex numbers, differential equations) — beyond what the first round actually requires. Students can solve such items only if they have acquired this knowledge outside regular classes, for instance through self-study or other enrichment activities. High scores may therefore also indicate a willingness to engage with STEM content independently — a disposition well suited to a competition round that requires months of self-directed work.

RIASEC-Conventional contributed negatively to predicted success. The respective items tap a preference for structured, routine, accuracy-focused tasks in scientific settings (e.g., entering measurements into databases). One plausible explanation is that students with this orientation are less inclined to invest the kind of effort the first-round demands: five months of voluntary, discretionary engagement with hard physics problems, where the right approach is rarely apparent at first sight, is a poor match for a preference for predictable, well-bounded work.

Physics and engineering-related skills also contributed notably to the prediction, consistent with the physics-focused content of the competition, and consistent with prior findings (Lind & Friege, 2001; Tschisgale, Steegh, Petersen, et al., 2024). That mathematics skills nonetheless carry more predictive signal than this more physics-specific indicator likely reflects the composition of the underlying instrument: the items combine physics with engineering content and primarily assess intuitive physics understanding rather than actual physics problem-solving ability required to successfully solve the competition tasks.

Finally, whether a student had skipped a grade contributed positively to predicted success — a notable result, given how few students skip a grade, that this indicator nonetheless emerged as a robust predictor. Grade-skipping is typically granted to students identified as cognitively advanced, and accelerated students consistently show above-average general cognitive ability in the literature on academic acceleration (Steenbergen-Hu et al., 2016). It may therefore capture some of the general cognitive ability that the domain-specific skill tests do not directly measure.

The absence of non-cognitive variables (except for RIASEC-Conventional) among the retained predictors does not contradict prior research identifying such variables as relevant for success in STEM competitions. Rather, this finding should be read in light of the distinction between explanatory and predictive modeling (Breiman, 2001b; Shmueli, 2010). Prior studies have primarily relied on explanatory models, which test whether a predictor makes a statistically detectable unique contribution after controlling for other variables; in such frameworks, self-related beliefs such as self-concept and expectancy of success have been shown to relate to STEM competition success (Steegh et al., 2021b; Tschisgale, Steegh, Petersen, et al., 2024). The elastic net applied here, by contrast, was designed for prediction and shrinks small contributions to zero in the interest of out-of-sample performance. Two things therefore follow. First, a non-cognitive variable can carry a genuine but modest association with success and still be dropped, because a small unique effect buys little predictive accuracy. Second, where non-cognitive and cognitive variables are correlated — as motivation and domain-specific skill tend to be — the penalty retains the stronger cognitive predictor and zeroes the non-cognitive one, so the latter's apparent irrelevance partly reflects shared variance rather than no association at all. Non-cognitive variables may thus remain meaningfully related to first-round success even though their unique predictive contribution, here, is small or even zero.

**Identifying Potentially Successful Non-Participants (RQ2)**

RQ2 asked to what extent potentially successful non-participants exist among non-participating students. To address this RQ, the final elastic net model trained on the participant data was applied to the non-participant data, and the subset of non-participants whose predicted probabilities of first-round success exceeded a precision-targeted decision threshold was identified. Although the retained model predicted individual success only modestly, its probability estimates were well-suited to the inference this study requires. The calibration curve departed from perfect calibration non-linearly, but only in ways immaterial here: identifying potentially successful non-participants draws exclusively on the upper tail of the calibration curve, above the decision threshold, where the model was, if anything, conservative (see Figure 1).

At the precision-targeted decision threshold of $p = .58$, 43 non-participants, corresponding to 3.9% of the non-participant sample, were identified as potentially successful. The small size of this subgroup should be interpreted in light of the conservative classification rule: Because precision was prioritized, the procedure identifies only those cases for whom the model provides comparatively certain predictions of likely first-round success.

The small size of this subgroup also reflects that the distribution of predicted success probabilities for the non-participant sample was shifted toward lower probabilities relative to the participant distribution (mean predicted success probability $\approx 36\%$ vs. $\approx 47\%$; see Figure 3). This shift is a direct consequence of the group differences in the model's cognitive predictors, on which participants exceed non-participants by more than one pooled standard deviation (see Table 1). Equally large group differences on the physics-specific motivational-affective variables and on physics and mathematics self-concept do not contribute to the shift, because the model assigns these variables zero or negligible weight.

Against the backdrop of prior evidence on predictors of participation, this pattern points to a divergence between the predictors of participation and those of success. Entry into STEM competitions is shaped primarily by motivational, social, and contextual factors (for an overview, see Steegh et al., 2019) — precisely the variables on which the two groups differ most strongly but which carry little predictive weight for success. Success once entered, by contrast, is mainly distinguished by cognitive skills, because participants are a self-selected group on the motivational variables, leaving domain-specific skills as what distinguishes those who advance from those who do not. This divergence is what creates the possibility of potentially successful non-participants — students whose cognitive profile is consistent with success once entered but whose motivational-affective profile is consistent with not participating in the first place. In these terms, the model captures only the second step: it estimates first-round success conditional on participation, not the decision to participate.

**Investigating Prior Competition Experience (RQ3)**

Examining the prior competition experiences of the potentially successful non-participants revealed that, as a group, they were already substantially engaged with the selective STEM competition system. Across nearly all selective competitions their prior participation exceeded that of the other non-participants, and in aggregate 74.4% had at least one prior selective entry, compared with 47.6% of the other non-participants. This pattern is consistent with the view that that the German STEM competition landscape, viewed as an ecosystem rather than as a set of isolated programs, reaches most potentially successful non-participants. Non-participation in the Physics Olympiad therefore does not necessarily place a student outside the broader system of STEM enrichment; the same student may participate in a mathematics competition, a computer science competition, or another STEM competition and still have access to opportunities for challenge, recognition, and support. Because this reach operates largely through other selective competitions, it likely reflects the comparatively dense German competition landscape; in systems with fewer such competitions, more potentially successful students might remain unreached.

A particularly telling contrast concerns the focal competition itself: 30.2% of the potentially successful non-participants had previously entered the Physics Olympiad, against only 3.7% of the other non-participants. A sizable share of these “non-participants” are thus former participants who did not enter in the focal year. For this part of the group, non-participation reflects intermittency rather than absence, and the relevant concern shifts from recruitment toward retention. More generally, non-participation in a specific year should not automatically be interpreted as problematic. Some students may have participated previously and found that this competitive, task-centered format is not well suited to their interests or preferred ways of engaging with STEM. From an ecosystem perspective, the relevant question is not whether every high-potential student participates continuously in the Physics Olympiad, but whether such students are reached by suitable enrichment somewhere within the broader landscape.

Against this backdrop, the unreached subgroup (11 potentially successful non-participants with no record of prior selective competition entry, roughly 1% of the non-participant sample) is small but substantively meaningful. These are the students whom the initiative-dependent selective routes never reached, and characterizing them (in particular, whether they cluster on background characteristics such as gender or socioeconomic status) will require larger samples and is left to future research.

**Implications**

The present findings suggest a shift in perspective on unrealized potential in STEM competitions: from the single competition to the broader landscape of related competitions. Where a competition is embedded in such a landscape, students who do not enter one may already be engaged through others. Reaching potentially successful students may therefore depend less on recruitment into any single competition than on the connections between competitions. In the present case, most potentially successful non-participants had already entered a selective competition (most often the Mathematics Olympiad), which points to concrete possibilities such as making participants in adjacent competitions aware of related opportunities, and, for the sizable share who had previously entered the Physics Olympiad itself, toward sustaining engagement across years rather than mainly recruiting anew. Such approaches presume, however, that non-participation reflects limited awareness; where it instead reflects an informed preference against the individual, task-centered format, raising awareness would be misdirected, so the appropriate course depends on why a given student did not enter—something the present data cannot determine.

Every measure described so far depends on an existing competition entry to work from — steering participating students toward adjacent programs, or bringing former participants back. The unreached potentials (roughly 1% of the non-participant sample) offer no such foothold, so these measures pass them by entirely. Whether a miss rate of this size is acceptable, or worth dedicated effort because these students are invisible to the initiative-dependent routes that currently structure entry, is not something the present data can decide; it depends on how a STEM competition weighs its mission against the resources reaching further would require, and is best left to organizers and governing bodies. Should organizers and governing bodies judge the unreached potentials worth pursuing, two gaps could explain the group. First, some schools may engage little with selective competitions at all, so no student in them is connected to the opportunity. Second, even in participating schools, teachers are typically the main point of contact through which students learn of selective competitions and are encouraged to enter (Abernathy & Vineyard, 2001; Tschisgale, 2024); where recruitment runs through their judgment, quietly able students may be overlooked. The two gaps call for different responses. The first is a matter of reach: reaching schools currently outside the system, whether through direct advertising to non-participating schools or through pre-service teacher education, since teachers who learn about the competition landscape at university carry that awareness into whatever schools they later join — including ones that do not yet engage. The second is a matter of how recruitment operates within participating schools, where students typically depend on a teacher to learn of a competition and be encouraged to enter; here the lever is strengthening teachers' capacity and inclination to identify and encourage suitable students. This study identifies the unreached potentials but cannot establish which gap produced them. These directions therefore remain conjectural rather than prescriptive.

Finally, the study carries a methodological implication for how missed potential in STEM competitions is assessed. Missed potential cannot be counted directly: the relevant population is unobserved, so one can only estimate, from a sample, the proportion of non-participants who are potentially successful — and, within these, the proportion the broader competition landscape leaves unreached. These two proportions must not be conflated. Reading the first as if it were the second, i.e. treating every potentially successful non-participant as missed by the system, overstates the gap, because many of these students prove to be already engaged through other competitions. A defensible estimate therefore requires examining the identified group's wider participation records, not just the size of that group. The two-step approach used here — model-based identification followed by this closer examination — offers a template for other, similarly structured competition systems, provided comparable data are available and the sample matches the target population in the respects the design controlled (here, school type and grade level).

**Limitations and Future Research**

First, although the trained machine learning model drew on a broad set of predictors, breadth does not guarantee that the most diagnostic indicators of first-round success were among them. Prior work has identified physics problem-solving ability as a central predictor of Physics Olympiad success (Tschisgale, Steegh, Petersen, et al., 2024), whereas the present feature set captured physics-related ability only through a physics and engineering-related skills instrument whose items combine physics with engineering content and primarily tap intuitive physics understanding rather than the actual problem-solving abilities the competition demands. Future work should incorporate measures more specific to problem solving and examine how far prediction of first-round success can be improved and to what extent prediction of success in higher competition rounds is possible.

Second, because non-participants have no observed competition outcome, the assumption that the participant-based predictive model transfers to non-participants cannot be validated within this study. The identified non-participant group should therefore be understood as a model-based estimate of who would likely have succeeded had they participated, not as a set of confirmed cases, and the proportion of potentially successful non-participants in particular should be read as approximate. This limitation is inherent to the counterfactual design rather than specific to the data, and cannot be resolved with cross-sectional data of the kind used here.

Third, although the overall non-participant sample was sizable ($N = 1{,}103$), the analysis narrowed to small groups: the model identified 43 potentially successful non-participants, and 11 of these had no prior selective competition experience. The size of the identified group is partly a consequence of the deliberately conservative decision threshold, but it nonetheless leaves the downstream quantities imprecise: the prior-participation proportions rest on 43 students, and the unreached share (about 1%) on just 11. These proportions should therefore be read as approximate rather than as precise population values. Because a larger non-participant pool would yield correspondingly larger groups, more precise estimates would require a substantially larger sample.

Fourth, the present findings are likely to generalize only to STEM competition systems that are structurally similar to the German one, in which a given competition is embedded in a dense landscape of adjacent selective competitions; the ecosystem interpretation in particular depends on this density, and in national systems organized differently—with fewer, less connected, or differently governed competitions—the balance between students reached through other pathways and those left entirely unreached is likely to differ. Comparable research across a range of countries is therefore needed before the present conclusions can be treated as broadly generalizable.

## CONCLUSION

STEM competitions are widely regarded as valuable enrichment activities that can positively shape participants' academic and career trajectories, and they are further intended to identify capable, science-interested students and support them — benefits that, on both counts, can be realized only for students who actually participate. Because the variables that predict participation differ from those that predict success, this study asked whether potentially successful students remain outside the German Physics Olympiad and, if so, whether they also lie beyond the reach of the broader selective-competition system. By training a calibrated machine learning model on participants and transferring its group-level predictions to a large non-participant sample, a small group of potentially successful non-participants was identified (43 students, 3.9%). The substantial majority of them, however, had already entered at least one other selective STEM competition (most often the German Mathematics Olympiad) leaving only about 1% of non-participants (11 students) with no prior participation in a selective STEM competition.

Viewed from the perspective of the German Physics Olympiad, these 43 students are unrealized potential. Viewed across the surrounding STEM competition ecosystem, however, most of them turn out to be already engaged elsewhere, leaving only a small group outside the system entirely. Whether counting the former as already reached is the appropriate reading depends on the goals a competition sets itself — in particular, whether prior engagement in an adjacent selective STEM competition is regarded as a comparable form of support — and this, like whether the small residual group represents an acceptable outcome, is a question for competition organizers and governing bodies rather than one the present study can settle. Independently of that judgment, the study establishes a methodological point: whether potential is genuinely missed is better assessed across a set of related competitions than within any single one — an ecosystem-level perspective that also offers a template for the same question in other competition systems.

## Declarations

***Ethics approval and consent to participate.*** The data used in this study was collected within the *WinnerS* project. Participation was voluntary and all ethics requirements for human subjects' research were met as testified by the ethics committee of the IPN under the approval number 2022_13_HO.

***Consent for publication.*** Not applicable.

***Competing interests.*** The authors declare that they have no competing interests.

***Use of AI tools.*** During the preparation of this work, the author used Anthropic's Claude to assist with language editing (grammar and spelling correction, and editing for clarity and style), with formulating and refining wording, and with drafting and debugging analysis code. All AI-generated or AI-assisted code was reviewed, tested, and verified by the author. The author reviewed all AI-assisted text and takes full responsibility for the content of the publication.

## Author Contribution Statement (CRediT)

Paul Tschisgale: Conceptualization, Formal Analysis, Methodology, Visualization, Writing – Original Draft Preparation, Writing – Review & Editing. Knut Neumann: Project Administration, Funding Acquisition, Writing – Review & Editing.

## Acknowledgments

The author is grateful to Jannis Zeller for helpful discussions on the preliminary data-analytic approach.

## Funding

This study used data from the *WinnerS* research project funded by the Leibniz Association (Germany) under Grant K194/2015.